\documentclass[prl,twocolumn,superscriptaddress]{revtex4}

\usepackage{psfrag}
\usepackage[T1]{fontenc}

\usepackage{amsfonts}
\usepackage{amssymb}

\usepackage{times}
\usepackage{pifont}
\usepackage{ae,aecompl}
\usepackage{amsmath}
\usepackage{graphicx}

\begin{document}

\title{Anomalous Scaling and Fractal Dimensions}
\author{Wolfhard Janke}
\affiliation{Institut f\"ur Theoretische Physik, Universit\"at Leipzig,
  Augustusplatz 10/11, 04109 Leipzig, Germany}
\author{Adriaan M. J. Schakel} \affiliation{Institut f\"ur Theoretische
Physik, Freie Universit\"at Berlin, Arnimallee 14, 14195 Berlin,
Germany}
\begin{abstract}
The relation between critical exponents, characterizing a continuous
phase transition, and the fractal structure of physical lines,
proliferating at the critical point, is established by considering the
two-dimensional O($N$) spin model for which many exact results are
available.
\end{abstract}

\date{\today}

\maketitle

A field theory describes a grand canonical ensemble of fluctuation loops
of arbitrary shape and length, a so-called \textit{loop gas}.  In the
context of quantum field theories, the fluctuating loops physically
represent the worldlines traced out by the quantum particles, while in a
classical context they represent, for example, high-temperature (HT)
graphs (in the case of spin models), current loops (in the case of the
Ginzburg-Landau theory) or closed stream lines (in the case of the
complex $|\phi|^4$ theory).  In a recent comment, Prokof'ev and
Svistunov~\cite{ProkofevSvistunov} pointed out that a relation proposed
by Hove, Mo, and Sudb{\o} \cite{Hoveetal} between the anomalous scaling
dimension of a field and the fractal or Hausdorff dimension of critical
loops is incorrect.  They concluded furthermore that the anomalous
scaling dimension cannot be deduced from simulations of closed loops
only.  They supported their criticism by Monte Carlo simulations of the
HT representation of the three-dimensional complex $|\phi|^4$ theory and
reported the value $D = 1.7655(20)$ as fractal dimension of the HT
graphs at the critical point.

As was first pointed out by Helfrich and M\"uller \cite{HM}, the HT
expansion of an $N$-vector spin or O($N$) lattice model, which can be
visualized by closed graphs along the bonds on the underlying lattice
\cite{Stanley}, describes at the same time a loop gas of sterically
interacting physical lines.  Specifically, the HT expansion of the
three-dimensional XY model ($N=2$), which belongs to the same
universality class as the $|\phi|^4$ theory, simultaneously represents
the vortex loop gas of a bulk superconductor \cite{GFCM}.  As the
magnetic interaction is screened, vortices in superconductors have
short-range interactions that can be accurately described by a steric
repulsion.  The equivalence implies in particular that the fractal
dimensions of HT graphs of the three-dimensional XY model and vortex
loops in bulk superconductors, studied in Ref.~\cite{Hoveetal},
coincide.  Prokof'ev and Svistunov~\cite{ProkofevSvistunov} therefore
rightly compare their results with those of Ref.~\cite{Hoveetal}, where
the value $D=1.92(4)$ was reported for magnetic vortex loops at the
critical point.

In their reply to the comment \cite{ProkofevSvistunov}, Hove and
Sudb{\o} \cite{HoveSudbo} questioned the validity of the criticism
leveled at their earlier work.  By considering the HT representation of
the two-dimensional O($N$) spin model for $-2 \leq N \leq 2$, for which
many exact results are available, we give in this short note analytic
arguments supporting the findings of Prokof'ev and Svistunov and,
hopefully, settle the issue.  The HT graphs of the two-dimensional Ising
model ($N=1$) physically represent Peierls domain walls, separating spin
clusters of opposite orientation on the dual lattice.  These closed
lines, which we recently investigated numerically in
Ref.~\cite{geoPotts}, form the two-dimensional counterpart of magnetic
vortex loops in three dimensions

A loop gas can be conveniently characterized by the loop distribution
$\ell_n$, giving the average number of loops of $n$ steps per unit
volume.  Close to the critical point $K_\mathrm{c}$, $\ell_n$
asymptotically takes a form similar to the cluster distribution in
percolation theory \cite{StauferAharony},
\begin{equation}
  \label{elln}
  \ell_n \sim n^{-\tau} \mathrm{e}^{- \theta n}, \quad \theta \propto
  (K-K_\mathrm{c})^{1/\sigma}.
\end{equation} 
Here, $\theta$ is the line tension, $\tau$ and $\sigma$ are two
exponents characterizing the distribution, and $K$ is the tuning
parameter. When the line tension is finite, the Boltzmann factor in the
distribution (\ref{elln}) exponentially suppresses large loops.  Upon
approaching the critical point, $\theta$ vanishes at a rate determined
by the exponent $\sigma$.  At $K_\mathrm{c}$, the loops proliferate as
they can now become arbitrary long without energy penalty.  The
remaining factor in the loop distribution is an entropy factor, giving a
measure of the number of ways a loop of $n$ steps can be embedded in the
lattice.  The entropy exponent $\tau$ is related to the fractal
dimension $D$ of the loops at the critical point via
\begin{equation} 
\tau=d/D+1,
\end{equation}  
where $d$ denotes the dimension of space (in the case of classical
theories) or spacetime (in the case of quantum theories).  The entropy
factor decreases with increasing $n$ because it becomes increasingly
more difficult for paths to close the longer they are.

In the limit $N\to 0$, the HT graphs of the O($N$) model reduce to
self-avoiding random walks (SAWs) which physically represent polymer
rings in good solvents \cite{deGennes}.  Using this equivalence, de
Gennes determined the fractal structure of SAWs in terms of the critical
exponents of the O($N\to 0$) model.  Generalizing his results to
arbitrary $-2 \leq N \leq 2$, we showed in a recent Letter \cite{ht}
that the fractal dimensions $D$ of the HT graphs and the exponents
$\sigma$ of the loop distribution (\ref{elln}) are related to the
correlation length exponent $\nu$ of the O($N$) model in the following
way:
\begin{equation}
  \label{nuD}
  \nu = 1/\sigma D.
\end{equation}
For SAWs ($N\to0$) $\sigma =1$, but in general $\sigma$ takes a
different value.

In the context of SAWs it is well established that loops alone do not
yield all the critical exponents of the universality class defined by
the O($N\to 0$) model \cite{deGennes}.  To that end, also the total
number $z_n \equiv \sum_{\mathbf{x}'} z_n(\mathbf{x}, \mathbf{x}')$ of
SAWs of $n$ steps starting at $\mathbf{x}$ and ending at an arbitrary
site $\mathbf{x}'$ is needed.  Because of translation symmetry
$z_n(\mathbf{x}, \mathbf{x}') = z_n(|\mathbf{x} - \mathbf{x}'|)$, and
$z_n$ does not depend on $\mathbf{x}$.  The ratio of $z_n(\mathbf{x},
\mathbf{x}')$ and $z_n$ defines the probability
$P_n(\mathbf{x},\mathbf{x}')$ of finding a path connecting $\mathbf{x}$
and $\mathbf{x}'$ in $n$ steps.  As $n \to \infty$, it scales as
\cite{Fisher}
\begin{equation}
  \label{P}
  P_n(\mathbf{x},\mathbf{x}') = z_n(\mathbf{x},\mathbf{x}')/z_n
  \sim n^{-d/D} \, \mathsf{P} \left( |\mathbf{x}-\mathbf{x}'|/n^{1/D}
  \right),
\end{equation} 
with $\mathsf{P}$ a scaling function, while the number $z_n$ scales as
\begin{equation}
\label{zn}
  z_n \sim n^{\vartheta/D} {\rm e}^{- \theta n},
\end{equation}   
with $\vartheta$ an exponent which is to be identified.  Since the
number of possible rooted open paths with no constraint on their
endpoint increases with the number  $n$ of steps, $\vartheta$ is expected
to be positive.

\begin{figure}
\centering
\psfrag{x}[t][t][1][0]{$\mathbf{x}$}
\includegraphics[width=0.3\textwidth]{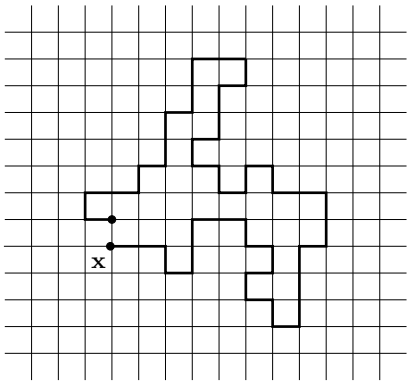}
\caption{A SAW in two dimensions returning to a site adjacent to its
starting point $\mathbf{x}$.
  \label{fig:saw}}
\end{figure}
The loop distribution (\ref{elln}) is related to the number of paths
$z_n(\mathbf{x},\mathbf{x}+ a \mathbf{i})=z_n(a)$ of $n$ steps returning
to a site $\mathbf{x}+ a \mathbf{i}$ adjacent to the starting point
$\mathbf{x}$ (see Fig.~\ref{fig:saw} and Ref.~\cite{deGennesbook})
through
\begin{equation}
  \label{ellz}
  \ell_n = \frac{1}{V} \sum_\mathbf{x} \frac{1}{n}  z_n(a),
\end{equation}  
with $\mathbf{i}$ any of the (positive or negative) directions on the
lattice, $V$ the lattice volume, and $a$ the lattice spacing.  The
latter serves as a microscopic cutoff, whence $z_n(a)$ rather than
$z_n(0)$ appears when closing open paths.  The factor $1/n$ in
Eq.~(\ref{ellz}) accounts for the fact that a loop can be traced out
starting at any lattice point along the loop.  As first shown by
McKenzie and Moore for self-avoiding random walks \cite{McKenzieMoore},
consistency of Eqs.~(\ref{ellz}) with $z_n(a) = z_n P_n(a)$ and
(\ref{elln}) requires that the scaling function $\mathsf{P}(t)$ must
vanish for $t \to 0$ and behave for \textit{small} argument $t$ as
$\mathsf{P}(t) \approx t^\vartheta$ with an exponent determined by the
\textit{asymptotic} behavior (\ref{zn}) of the number $z_n$ of open
paths at the critical point ($\theta=0$).

The fractal dimension $D$ and the exponents $\sigma$ and $\vartheta$
determine the critical exponents of the theory.  As for self-avoiding
random walks, the relations can be derived by writing the correlation
function $G(\mathbf{x},\mathbf{x}')$ in terms of
$z_n(\mathbf{x},\mathbf{x}')$ as $G(\mathbf{x},\mathbf{x}') \sim \sum_n
z_n (\mathbf{x},\mathbf{x}')$, where because of translational
invariance, $G(\mathbf{x}, \mathbf{x}') = G(|\mathbf{x}- \mathbf{x}'|)$.
When evaluated at the critical point, where $G(\mathbf{x},\mathbf{x}')
\sim 1/|\mathbf{x}-\mathbf{x}'|^{d-2 + \eta}$, this gives the relation
proposed by Prokof'ev and Svistunov~\cite{ProkofevSvistunov}
\begin{equation} 
\label{vartheta}
\eta = 2 - D - \vartheta.
\end{equation}
Using high-precision data for $\eta$, together with their estimate for
$D$, they arrived at the estimate $\vartheta = 0.1965(20)$ for the
three-dimensional $|\phi|^4$ theory.  The relation proposed by Hove, Mo,
and Sudb{\o} \cite{Hoveetal} corresponds to setting $\vartheta$ to zero
in Eq.~(\ref{vartheta}), which in general is not allowed.  Given the
exact values for $\eta$ \cite{Nienhuis} and the fractal dimensions $D$
of the HT graphs \cite{Vanderzande}, $\vartheta$ can be determined
exactly for the two-dimensional O($N$) model (see Table~\ref{table:On}).
Through the exact enumeration and analysis of the number $z_n$ of
self-avoiding walks on a square lattice up to length 71, the expected
value $\vartheta/D=11/32$ for $N=0$ has been established to high
precision \cite{Jensen04}, while our Monte Carlo simulation
\cite{geoPotts} of the high-temperature representation of the
two-dimensional Ising model unambiguously shows that $\vartheta$ is
nonzero.  In their numerical study of the three-dimensional $|\phi|^4$
theory \cite{ProkofevSvistunov}, Prokof'ev and Svistunov considered not
the asymptotic behavior of the number $z_n$ of open graphs to verify the
value $\vartheta = 0.1965(20)$, but the behavior of the probability
(\ref{P}) for small arguments.  Both should give the same results,
according to our analysis.
\begin{table}
  \begin{tabular}{l|r|ccc|ccc}
%
%
    Model & $N$ & $\gamma$ & $\eta$ & $\nu$ & $D$ & $\sigma$ &
    $\vartheta$ \\[.05cm]
    \hline & & & & & & \\[-.4cm]
    Gaussian & $-2$ & $1$ & $0$ & $\frac{1}{2}$ & $\frac{5}{4}$ &
    $\frac{8}{5}$ & $\frac{3}{4}$\\[.1cm]
    SAW & $0$ & $\frac{43}{32}$ & $\frac{5}{24}$ & $\frac{3}{4}$ &
    $\frac{4}{3}$ & $1$ & $\frac{11}{24}$ \\[.1cm]
    Ising & $1$ & $\frac{7}{4}$ & $\frac{1}{4}$ & $1$ & $\frac{11}{8}$ &
    $\frac{8}{11}$ & $\frac{3}{8}$ \\[.1cm]
    XY & $2$ & $\infty$ & $\frac{1}{4}$ & $\infty$ & $\frac{3}{2}$ & $0$
    & $\frac{1}{4}$ \\[.1cm]

    Spherical & $\infty$ & $\infty$ & $0$ & $\infty$ & $2$ & $0$
    & $0$ 
%
%
  \end{tabular}
  \caption{Critical exponents of the two-dimensional critical O($N$)
    spin models, with $N=-2,0,1,2,\infty$, respectively, together with the the
    fractal dimension $D$ of the HT graphs as well as the two exponents
    $\sigma$ and $\vartheta$.
    \label{table:On}}
\end{table}

Relation (\ref{nuD}) can, incidentally, be derived by using the
second-moment definition of the correlation length $\xi$,
\begin{equation} 
  \xi^2 = \frac{\sum_\mathbf{x} x^2 G(x)}{\sum_\mathbf{x} G(x)},
\end{equation} 
with $x \equiv |\mathbf{x}|$.  Finally, using the definition of the
susceptibility $\chi$, $\chi = \sum_{\mathbf{x}'}
G(\mathbf{x},\mathbf{x}')$, which diverges as $\chi \sim
|K-K_\mathrm{c}|^{-\gamma}$, we obtain
\begin{equation} 
\label{gamma}
\gamma = (D + \vartheta)/\sigma D .
\end{equation} 
This relation generalizes one due to Cloizeaux \cite{Cloizeaux} for SAWs
for which $\sigma=1$.  The explicit expressions for $\nu,\eta$, and
$\gamma$ satisfy Fisher's scaling relation, $\gamma/\nu = 2-\eta$.  The
critical exponents are a function of the two independent variables $z_1
\equiv D+\vartheta$ and $z_2 \equiv\sigma D$ which determine the
anomalous scaling dimension of the $\phi$ and $\phi^2$ fields:
\begin{eqnarray} 
d_\phi &=& \tfrac{1}{2} ( d -2 - \eta) = \tfrac{1}{2} ( d - z_1),
\nonumber \\ \quad d_{\phi^2} &=& d - \frac{1}{\nu} = d - z_2,
\end{eqnarray} 
respectively.

With $\vartheta$ set to zero, Eq.~(\ref{gamma}) reduces to the relation
$\gamma = 1/\sigma$ proposed by Nguyen and Sudb{\o} in
Ref.~\cite{NguyenSudbo}.  Given that $\vartheta$ is in general positive,
that relation, too, is generally incorrect. It is, however, not
impossible for $\vartheta$ to be zero.  A class of models with the
special value $\vartheta=0$ is provided by a \textit{noninteracting}
Bose gas in $d$ space dimensions with modified energy spectrum
$\epsilon(k) \propto k^D$ with $D< d \leq 2D$ at fixed particle number
density \cite{pre01}.  This class belongs to the universality class of
the spherical model, obtained by taking $N \to \infty$ in the O($N$)
model, with long-range interactions \cite{GuntonBuckingham}.  The
critical exponents of this universality class are exactly known
\cite{Joyce}.  Those of the two-dimensional spherical model without
long-range interactions, corresponding to $D=2$ and zero critical
temperature, are included in Table~\ref{table:On}.

The presence of the exponent $\vartheta$ is specific to having
geometrical objects, viz.~lines, that can be open or closed as it
provides consistency when closing open lines.  For geometrical objects
where the notion of open and closed does not apply, such as for spin
clusters, the analog of the exponent $\vartheta$ is absent.  This is,
for example, the case in the Fortuin-Kasteleyn formulation of the
$Q$-state Potts model \cite{FK}, and for the improved estimators in the
cluster update Monte Carlo algorithms \cite{SwendsenWangWolff}.

In conclusion, using exact results for the two-dimensional O($N$) model
with $-2 \leq N \leq 2$, we provided analytic support to the findings of
Ref.~\cite{ProkofevSvistunov} that the full set of critical exponents
for a given theory cannot be determined from the loop distribution
alone.  For the full set also the asymptotic behavior of the number of
\textit{open} paths is needed.  Generalizing results known for SAWs, we
showed that the exponent $\vartheta$ governing this behavior is
connected to that of the probability $P_n(\mathbf{x},\mathbf{x}')$ of
finding a path connecting $\mathbf{x}$ and $\mathbf{x}'$ in $n$ steps
for $\mathbf{x}' \to \mathbf{x}$, i.e., in the limit of closing  open
paths.  Finally, we showed that this connection, together with a
relation recently proposed by us \cite{ht}, determines the full set of
critical exponents in terms of the fractal structure of open and closed
physical lines.

\end{document}